# Catheter Monitoring in Intelligent Endovascular Navigation Systems: Interactive Simulations and Mixed Reality for Enhanced Navigational Awareness

**Authors:** Veronica Ruozzi[1,*], Giovanni Battista Regazzo[2], Maria Chiara Palumbo[1], Wim-Alexander Beckers[2], Mouloud Ourak[2], Xiu Zhang[1], Francesca Perico[1], Alessandro Caimi[1], Emmanuel Vander Poorten[2], Emiliano Votta[1]

**Affiliations:**
[1]Department of Electronics, Information and Bioengineering, Politecnico di Milano, Italy
[2]Department of Mechanical Engineering, KU Leuven, Belgium

**Corresponding author:** Veronica Ruozzi, `veronica.ruozzi@polimi.it`

## Abstract

**Purpose** Developing and testing a framework that integrates real-time catheter shape reconstruction, interactive simulations, and mixed reality visualization to enable accurate monitoring of catheter–vessel interactions during endovascular navigation.

**Methods** A finite element model (FEM) of the venous pathway from the right femoral vein to the inferior vena cava was generated from computed tomography data and implemented into an interactive simulation. Catheter motion was imposed as boundary condition, and catheter–vessel contact was modeled with a Lagrange multiplier formulation to compute vessel deformation. The framework was integrated and tested in-vitro using a sensorized catheter with Fiber Bragg Grating and electromagnetic sensors to provide real-time 3D shape and location as it was advanced by a catheter driver through a silicone replica of the vascular anatomy. Upon registration, real-time sensor read-outs fed the simulation, and the updated catheter and vessel geometries were streamed to Hololens 2$^{TM}$ (HL2). The performance of the simulation and the accuracy of FEM-computed vessel wall displacement were validated vs. experimental ground-truth obtained via stereo frames triangulation.

**Results** Complexity and extent of catheter–vessel interaction affected FEM performance by increasing the computational cost. The simulated time exceeded the real temporal extent of the physical phenomenon by 12% during the initial navigation phase and by 45% when the catheter reached the most tortuous portion of the vessel. The HL2 rendering remained stable between 35 and 40 frames per second. Across these two phases, the median *relative displacement error* between FEM-computed vessel wall displacements and the ground-truth remained below 1 mm and 2.33 mm, respectively.

**Conclusion** The study demonstrates the feasibility of integrating interactive biomechanical simulation with real-time sensor data to enable continuous monitoring of catheter–vessel interactions, with mixed reality visualization serving as a user interface to support a more engaged and better-informed operator throughout the navigation.

# 1 Introduction

Transfemoral access is widely used for transcatheter procedures, from structural cardiac interventions to electrophysiology. Slender catheters are advanced through the vasculature to deploy implantable devices. Although these approaches offer major advantages over open surgery [1], they remain technically demanding and require substantial operator expertise [2]. A key limitation is the lack of direct visualization of catheter–vessel interaction, as guidance relies mainly on X-ray fluoroscopy. This exposes patients and operators to radiation, provides only 2D projections, and yields transient lumen visualization limited by contrast washout, resulting in incomplete information. Consequently, catheter–vessel interactions and potential damage are poorly monitored, and vascular complications remain a concern, particularly in tortuous or calcified vessels or when large or stiff catheters are used [3, 4].

Although physics-based dynamic modeling approaches have been explored in laparoscopic settings [5], their application to endovascular navigation remains largely unexplored, mainly due to the high computational cost inherent to accurately modeling catheter–vessel contact interactions, which limits interactivity. Current head-mounted display (HMD)-based solutions also lack the accuracy and reliability required for manual tasks [6] and surgical guidance [7], while clinical evidence of clear benefit remains limited [8].

A recent endovascular-navigation study shows that combining a gamepad-based catheter controller with an HMD improves performance and navigation intuitiveness over standard displays [9]. These findings suggest that, although not yet accurate enough for standalone guidance, MARS can serve as effective interfaces that enhance operator understanding when integrated with more advanced surgical-workflow systems [9–11]. Recent advances in robotics and automation still rely on the surgeon's final decision, assuming full situational awareness. In this context, integrating MARS is promising [12, 13], as overlaying real-time, computer-generated information onto the surgical scene without obstructing the field can actively support intraoperative decision-making [8].

On this basis, this study introduces a method that integrates sensor-based catheter shape reconstruction, interactive biomechanical simulation, and mixed reality visualization into a MARS, enabling 3D monitoring of catheter–vessel interaction and associated vessel deformation, and reports its in-vitro integration and validation.

# 2 Methods

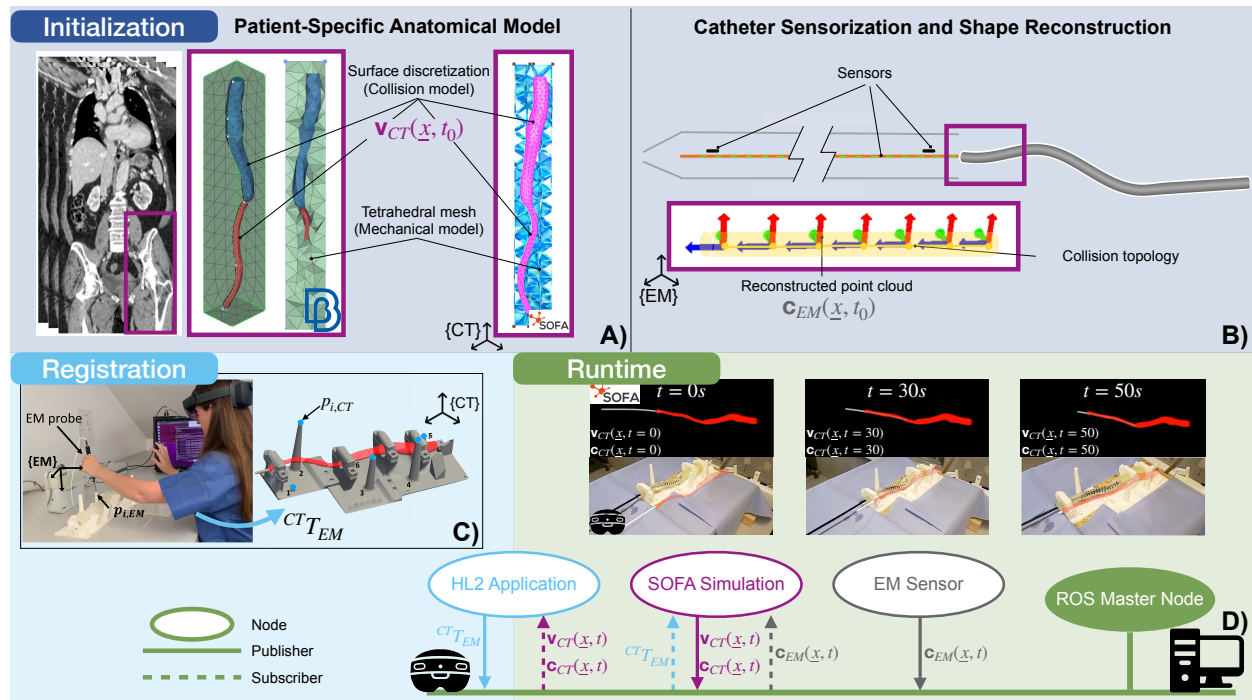

Figure 1: **Schematic overview of the proposed framework**. An *Initialization* step preprocesses vessel (A) and catheter (B) models and instantiates the interactive simulation. A *Registration* step aligns sensor ($\{EM\}$) and anatomical ($\{CT\}$) frames (C). During *Runtime*, the interactive simulation continuously updates the mixed reality visualization (D).

The method integrates four components: (i) an fiber Bragg grating (FBG)- and electromagnetic (EM)-instrumented catheter for real-time shape reconstruction; (ii) a finite element model (FEM) of the vessels for interactive deformation; (iii) hardware for sensor readout and frame alignment; and (iv) an mixed reality (MR) interface catheter monitoring. These elements operate within a three-step workflow, *Initialization*, *Registration*, and *Runtime*, illustrated in Fig. 1. The following subsections describe the approach using in-vitro experiments.

## 2.1 In-Vitro Setup

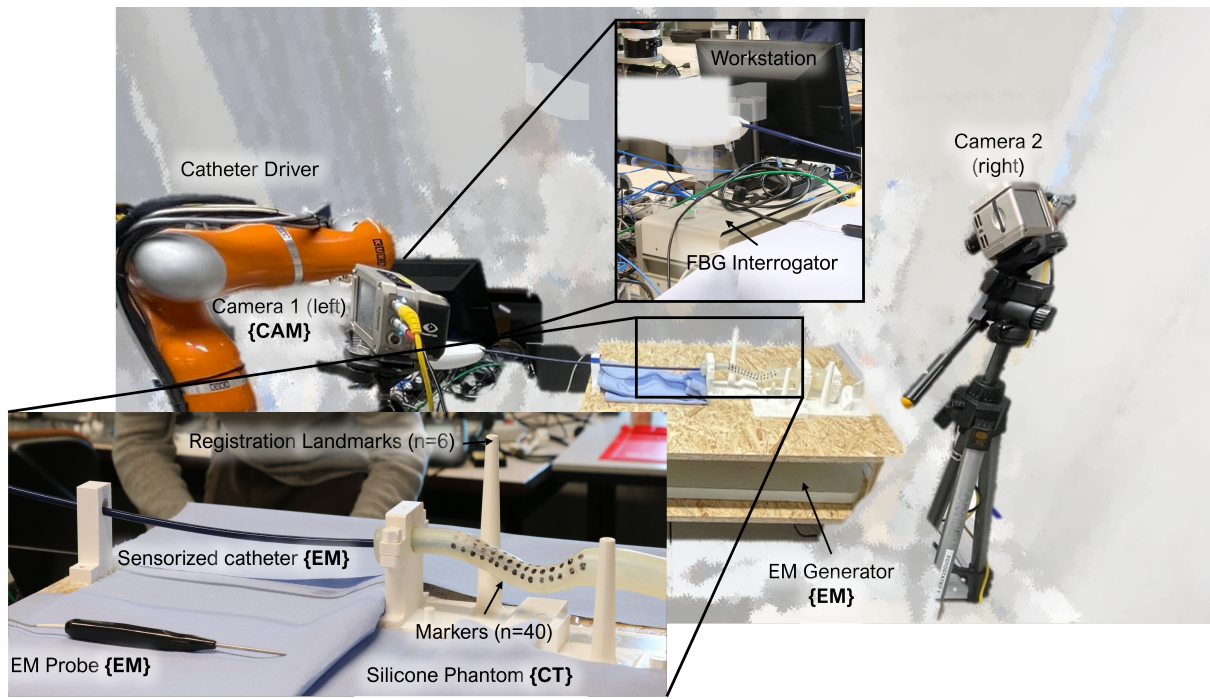

Figure 2: **Experimental setup for integrating and testing the framework.** A robotic driver inserted the sensorized catheter into a silicone phantom replicating the vessel anatomy. Two high-speed cameras tracked surface-marker displacements on the phantom wall, while the corresponding FEM nodes were identified by applying the transformation $^{CT}\mathbf{T}_{EM}$ to the point cloud acquired at rest using the EM probe.

The setup (Fig. 2) used a silicone phantom of the venous lumen from the femoral vein to the inferior vena cava (45.6 cm), derived from contrast-enhanced CT scans. Veins were cast with 4.5 mm walls, mounted on a 3D-printed support with six landmarks, and marked with 40 surface points for deformation tracking. A 7.4 mm catheter diameter was instrumented with two 5 DOF EM sensors (NDI - Northern Digital Inc., ON, Canada) and a 2 m-long multicore FBG sensor (38 gratings, 10 mm spacing; FBGS, Geel, Belgium). EM sensors were placed at the tip and 18 cm proximally [14]. Shape was reconstructed in real time (40 Hz) using twist-compensated calibration and the helical extension method [15] for 3D reconstruction, fused with EM data for absolute positioning and resampled at 78 points ($\mathbf{c}_{EM}(\underline{x}, t) = \{\mathbf{c}_{iEM}(t)\}_{i=1}^{78}$) over 37 cm. Additional components included a robotic driver [16], FBG interrogator, EM generator, EM Aurora® NDI Probe (Aurora, Northern Digital Inc., Waterloo, ON, Canada), Hololens 2™ (HL2) (Microsoft Inc.) HMD, and two high-speed stereo cameras (Phantom Miro 3, Vision Research Inc., Wayne, NJ, USA). Cameras tracked marker displacement, while the Aurora probe provided reference points for registration.

## 2.2 Initialization

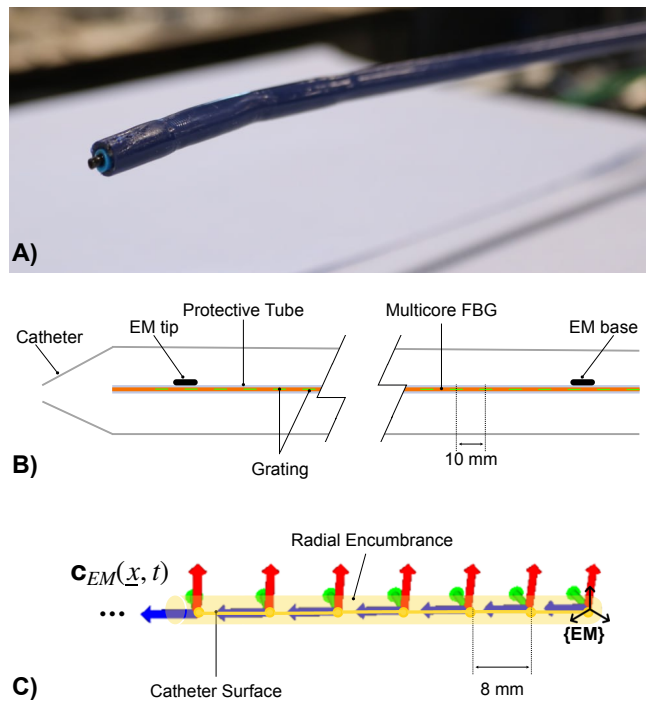

Figure 3: A 7.4 mm catheter (**A**) housed a multi-core FBG fiber and two EM sensors within its central lumen (**B**). The sensor-based reconstructed catheter centerline, expressed in the $\{EM\}$ reference frame as $\mathbf{c}_{\text{EM}}(\underline{x}, t)$, was uniformly discretized into 78 points. Each point served as the center of a rigid sphere, and consecutive points were connected by cylinders of 7.4 mm diameter, producing a piecewise-continuous geometric avatar of the catheter outer surface for collision and contact handling (**C**).

Virtual avatars of the CT-derived vein (Fig. 1A) and catheter (Fig. 1B) were generated to initialize the FEM.

**Vein** The segmented lumen was imported into ANSA (Beta CAE Systems, USA) and embedded in a hexahedral bounding box (*BBox*) expanded by 10% along the main vein axis (Z) and 20% along the others (X and Y) of the anatomical reference frame (i.e., $\{CT\}$). The lumen and outer *BBox* surfaces were remeshed with triangular elements using characteristic lengths of 8 mm (femoral–iliac segment), 15 mm (inferior vena cava), and 30 mm (*BBox*). The bulk of the equivalent continuum was discretized into linear tetrahedral elements (maximum aspect ratio 5). The *BBox* modeled surrounding tissue constraints as a deformable continuum with density $\rho = 1000 \ kg/m^3$ and isotropic linear elastic properties (Young's Modulus ($E$)=1 MPa, Poisson's Ratio ($\nu$)=0.36). This reduced the number of solid elements while maintaining bending and torsional deformability and introduced a deformable constraint simulating soft-tissue containment, limiting edge effects caused by node constraints. Mesh size, mechanical parameters, and discretization time step ($dt$) were chosen from sensitivity analyses in Supplementary Section S1.

**Catheter** The discretized centerline of the catheter avatar consisted in the set of 78

points $\mathbf{c}_{\text{EM}}(\underline{x}, t) = \{\mathbf{c}_i(t)\}_{i=1}^{78}$ whose coordinates were in the electromagnetic frame (i.e., $\{EM\}$) and continuously monitored via EM and FBG sensors. Consistently with the cross-sectional dimensions of the real catheter, a rigid sphere with 7.4 mm diameter was centered at each point $\mathbf{c}_i(t)$, and a cylinder with a 7.4 mm cross-sectional diameter was aligned with the segment running between consecutive points $\mathbf{c}_i(t)$ and $\mathbf{c}_{i+1}(t)$, with $i = 1, ..., 77$.

## 2.3 Registration

The rigid transformation matrix $^{CT}\mathbf{T}_{EM}$ aligning the $\{EM\}$ to the $\{CT\}$ reference frame was computed via a landmark-based Singular Value Decomposition (SVD) registration. Six pillars were integrated into the phantom support (Fig. 2) around the anatomical region of interest to span the EM field generator's working volume across different heights and served as registration landmarks. Their spatial coordinates in $\{CT\}$ ($\mathbf{p}_{i,CT}$, $i = 1, ..., 6$) were known a priori, while corresponding points in $\{EM\}$ ($\mathbf{p}_{i,EM}$, $i = 1, ..., 6$) were acquired by placing the EM probe into dedicated holders at the top of each pillar (Fig. 1C). The resulting registration yielded a Root Mean Square Error (RMSE) of 0.982 mm. Registration was managed through a MR interface developed in Unity3D (Unity Technologies, version 2020.3.25LTS) with the Mixed Reality Toolkit (MRTK v2.7.3), deployed to HL2.

## 2.4 Runtime

After initialization and registration, catheter point coordinates in the $\{CT\}$ frame ($\mathbf{c}_{CT}(\underline{x}, t)$) were streamed to the SOFA framework [17], with

$$\mathbf{c}_{CT}(\underline{x}, t) = {}^{CT}\mathbf{T}_{EM}\, \mathbf{c}_{EM}(\underline{x}, t).$$

Vessel–catheter interaction was modeled using a corotational FEM formulation with a Lagrange multiplier (LM)-based collision model applied to both lumen and catheter surfaces. The LM method provides physically accurate, parameter-free contact forces and strong numerical robustness, but requires a direct $LDL^T$ factorization of the linear system at each $dt$ [18]. Computational details are in Supplementary Section S2, with time-integration information in Section S1. Two constraint sections were defined at the proximal and distal *BBox* faces. For visualization, a 2 mm triangular mesh was mapped to the lumen surface, and its vertex coordinates ($\mathbf{v}_{CT}(\underline{x}, t)$) were streamed at each $dt$ to the MR application for updating.

Data transfer from sensors to simulation and from simulation to the MR application was handled via robotic operating system (ROS) inter-node communication (Fig. 1D).

## 2.5 In-vitro Experiments

The experiment was repeated three times. After registration, the EM Aurora® probe recorded the initial positions of all outer-surface markers, generating the reference point cloud $\mathbf{p}_{EM}(\underline{x}_i, t_0)$ with $i = 0, \ldots, 39$. Three points per clamp on the rigid 3D-printed base were also recorded with the probe to update the FEM constraint sections. The stereo cameras were calibrated using a checkerboard pattern. The robotic driver then advanced the catheter into the silicone phantom at $5, \text{mm/s}$ without active steering, a speed selected based on prior work [16] and commercial robotic catheter driver capabilities.

The start of catheter insertion triggered the synchronized acquisition of two datasets:

1. *experimental dataset*, stereo image sequences from the left and right cameras (Phantom Camera Control software, Vision Research Inc., USA);

2. *computational dataset*, i) a *rosbag* logging the 78 catheter points at 40 Hz vs. physical time; ii) an ASCII file containing the nodal coordinates of the catheter and vessel avatars as a function of the central processing unit (CPU) computation time (accumulated machine time ($T_{CPU}$), where $T_{CPU}$ denotes the time required by the CPU to compute and update the current mesh state.

At the end of each experiment, 13 time-points spanning catheter motion were selected, with 8 s between consecutive points. For each time-point, we quantified:

1. interactive performance – comparing $T_{CPU}$ from ASCII files with physical time ($T_{PHYS}$) from *rosbags*, using the one-to-one correspondence between real and simulated catheter points; HL2 rendering frames per second (FPS) was also monitored.

2. the accuracy of the simulated vessel deformation – markers on the outer surface of the silicone phantom were manually tracked in stereo images and reconstructed in 3D via triangulation [19], yielding $\mathbf{p}_{CAM}(\underline{x}, t)$. At $t = t_0$, a principal component analysis-based registration computed $^{EM}\mathbf{T}_{CAM}$ aligning $\mathbf{p}_{CAM}(\underline{x}, t_0)$ to $\mathbf{p}_{EM}(\underline{x}, t_0)$. This transformation was then composed with the transformation matrix $^{CT}\mathbf{T}_{EM}$ to express $\mathbf{p}_{CAM}(\underline{x}, t)$ in the $\{CT\}$ reference frame:

$$\mathbf{p}_{EXP}(\underline{x}, t) = {^{CT}}\mathbf{T}_{EM} \cdot {^{EM}}\mathbf{T}_{CAM} \cdot \mathbf{p}_{CAM}(\underline{x}, t)$$

Marker ground-truth displacements were computed from $\mathbf{p}_{EXP}(\underline{x}, t)$. At $t = t_0$, each marker was associated with its nearest neighbor $\mathbf{p}_{COMP}(\underline{x}, t_0)$ in the simulated lumen mesh $\mathbf{v}_{CT}(\underline{x}, t_0)$. The mismatch between the time-dependent 3D coordinates $\mathbf{p}_{COMP}(\underline{x}, t)$ computed by SOFA and the ground-truth time-dependent coordinates $\mathbf{p}_{EXP}(\underline{x}, t)$ was quantified as:

$$d(\underline{x}, t) = \|\mathbf{p}_{COMP}(\underline{x}, t) - \mathbf{p}_{EXP}(\underline{x}, t)\|$$

To rule out the effects of initial registration errors, a further metric, the *relative displacement error*, was defined as:

$$\Delta d(\underline{x}, t) = |d(\underline{x}, t) - d(\underline{x}, t_0)|$$

## 3 Result

### 3.1 Interactive Performance

The comparison of $T_{CPU}$ (SOFA) vs. $T_{PHYS}$ (*rosbag*) highlighted three phases in the experiments (Table 1, Fig. 4):

1. *Startup latency* – an initial lag by $4.66 - 6.94$s was observed.

2. *Low-latency phase*, until the catheter tip reached the most tortuous tract of the vein phantom – this phase required $63.83-68.30$s, $T_{CPU}$ progressed slightly more slowly than $T_{PHYS}$. The linear regression of $T_{CPU}$ on $T_{PHYS}$ data yielded $T_{CPU} = 1.12T_{PHYS}+2.10$ ($R^2 = 0.999$).

3. *High-latency phase*, beyond $T_{PHYS}$=56.00s – the difference between $T_{CPU}$ and $T_{PHYS}$ markedly increased, with $T_{CPU} = 1.45T_{PHYS} - 18.70$ ($R^2 = 0.993$).

Table 1: Cumulative physical time ($T_{PHYS}$), simulated time ($T_{CPU}$), and lag at sampled time-points.

| Time-point ID | $T_{PHYS}$ [s] | $T_{CPU}$ [s] | Lag [s] |
| --- | --- | --- | --- |
| 0 | 0.00 | [4.66–6.94] | [4.66–6.94] |
| 1 | 8.00 | [9.13–13.27] | [1.13–5.27] |
| 2 | 16.00 | [19.43–22.73] | [3.43–6.73] |
| 3 | 24.00 | [28.78–32.34] | [4.78–8.24] |
| 4 | 32.00 | [38.35–41.40] | [6.35–9.40] |
| 5 | 40.00 | [46.97–49.73] | [6.97–9.73] |
| 6 | 48.00 | [56.34–57.93] | [8.34–9.93] |
| 7 | 56.00 | [63.83–68.30] | [7.83–12.30] |
| 8 | 64.00 | [74.04–77.57] | [10.04–13.57] |
| 9 | 72.00 | [82.23–86.75] | [10.23–14.75] |
| 10 | 80.00 | [93.19–97.93] | [13.19–17.93] |
| 11 | 88.00 | [106.74–109.48] | [18.74–21.48] |
| 12 | 96.00 | [121.41–123.67] | [25.41–27.67] |

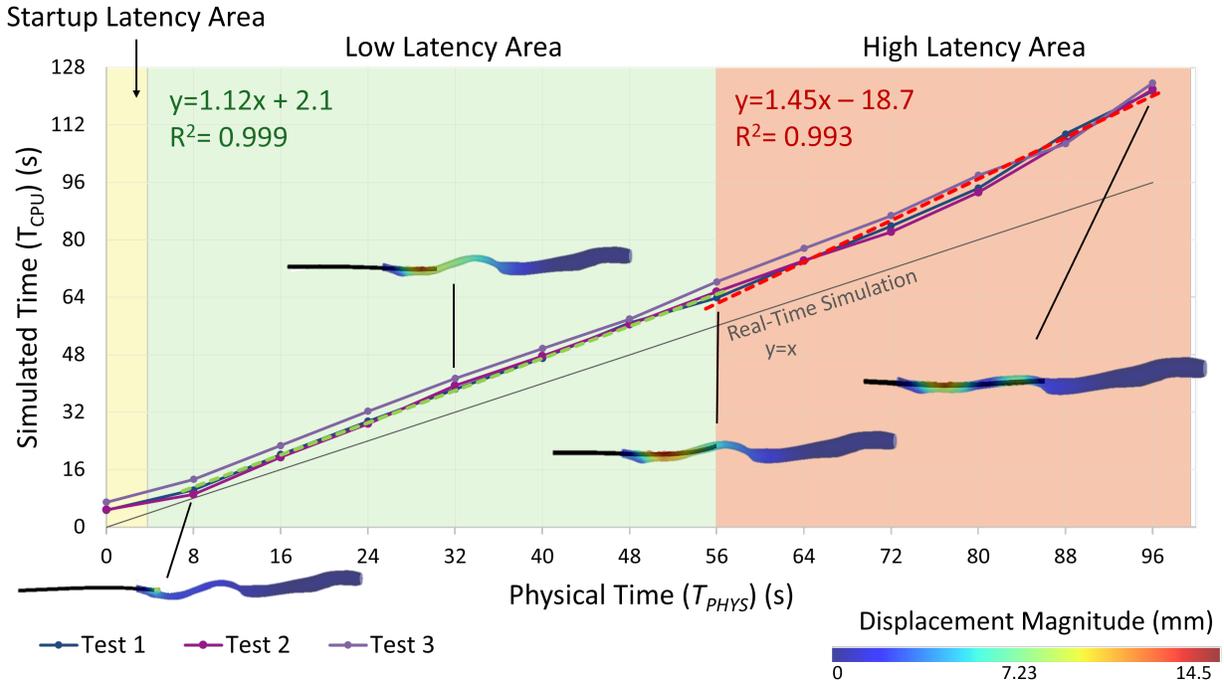

Figure 4: **Interactive FEM performance**: simulated time ($T_{CPU}$) vs. physical time ($T_{PHYS}$) for three repeated tests. Three latency phases were identified: startup, low-latency (slope 1.12), and high-latency (slope 1.45). The deformed vein lumen, color-coded by displacement magnitude, is shown at specific time-points.

The low variability in the initial lag suggests that startup latency is a fixed delay from system initialization and communication protocols, creating a brief dead zone before data reach the simulation. Lag accumulation during the low- and high-latency phases is mainly due to the increasing computational cost of growing contact interactions. In the low-latency phase, lag grows moderately (slope 1.12 in the $T_{CPU}$ vs. $T_{PHYS}$ regression), as catheter-lumen contact involves few mesh nodes. In the high-latency phase, the expansion of the contact region leads to faster lag accumulation, reflected by the higher slope (1.45; Fig. 4).

Throughout the experiment, the HL2 rendering FPS remained stable at 35-40 Hz.

## 3.2 Accuracy of Simulated Vessel Deformation

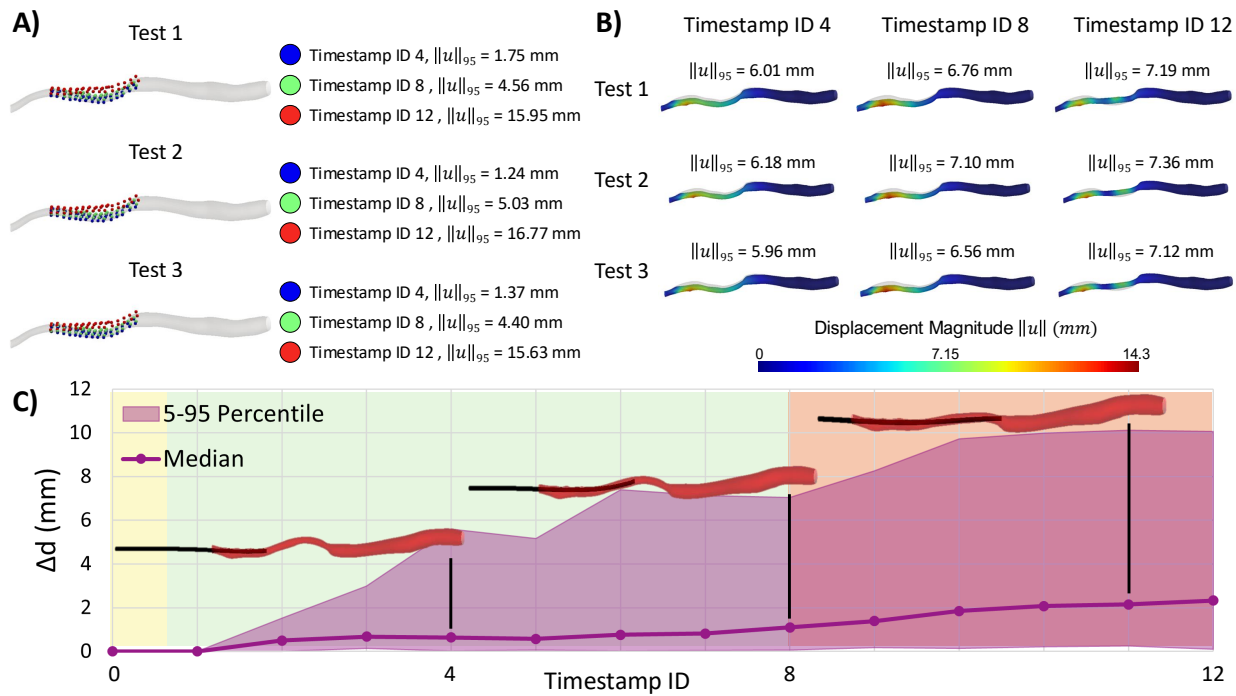

Figure 5: **A)** Ground-truth marker configurations ($\mathbf{p}_{EXP}$) overlaid on the undeformed vessel (light gray) at three representative time-points; each is annotated with the 95th percentile of the displacement magnitude ($\|\boldsymbol{u}\|_{95}$). **B)** Corresponding FEM configurations at the same time-points, shown with a displacement colormap and overlaid on the undeformed vessel. The $\|\boldsymbol{u}\|_{95}$ computed over the marker-covered region is reported. **C)** *Relative displacement error* for the three repeated tests, expressed as $\Delta d(\underline{x}_i, t_j)$. Median errors across the 13 time points are shown, with shaded bands indicating the 5th–95th percentile range. Representative FEM vessel–catheter configurations are shown at the three time-points.

Fig. 5A shows that the low- and high-latency phases correspond to markedly different vessel deformations, as quantified by the 95th percentile of marker displacement magnitude ($\|\boldsymbol{u}\|_{95}$). During the low-latency phase, $\|\boldsymbol{u}\|_{95}$ remained below 2 mm until the catheter tip reached the most tortuous vessel tract (time-point ID 4), increasing to about 5 mm when the tip was halfway through this tract (time-point ID 8) and the catheter has not yet straightened the vessel phantom. At the end of the high-latency phase (time-point ID 12), $\|\boldsymbol{u}\|_{95}$ reached nearly 17 mm, corresponding to a clear straightening of the initially tortuous tract.

This straightening behavior was well captured by the FEM model (Fig. 5B). However, the $\|\boldsymbol{u}\|_{95}$ computed over the lumen region covered by markers underestimated the global deformation, yielding values between 5.96 and 7.36 mm across the three-time points, consistently across tests.

A more reliable comparison is provided by the *relative displacement error* (Fig. 5C), whose median value remained below 1 mm throughout the early insertion phase, and increased progressively up to 2.33 mm as the catheter advanced through the most tortuous region, where lumen displacements were largest.

The uncertainty in reconstructing $\mathbf{p}_{EXP}(\underline{x},t)$ arises from stereo calibration re-projection error (0.784 mm) and from the estimation of $^{EM}\mathbf{T}_{CAM}$ and $^{CT}\mathbf{T}_{EM}$ (0.700 mm and 1.050 mm RMSE, respectively), resulting in a cumulative uncertainty of 2.534 mm. Accordingly, the median *relative displacement error* remained within reconstruction uncertainty throughout insertion. However, as catheter–vessel interaction increased in highly curved regions, the error distribution widened, with the 95[th] percentile reaching 9.97 mm in the narrowest and most tortuous segment (Fig. 5C).

## 4 Discussion

### 4.1 Novelty

While demonstrating interactive feasibility for robotic navigation, the proposed approach introduces two design choices:

- Vessel modeling – Unlike prior work [20], which modeled vessels as isolated thick-walled conduits, we embed the lumen in a deformable continuum. This incorporates surrounding-tissue constraints and removes the need to infer wall thickness from imaging. Leveraging intra-procedural US, future work could enable a pipeline where, after CT-US registration [21, 22], the continuum is assigned patient-specific $(E, \nu)$ via non-invasive vessel compliance and distensibility estimation [23].

- Catheter representation – The distal configuration of a sensorized catheter is reconstructed and imposed as a kinematic boundary condition, removing shape uncertainty and enabling interactive simulation by reducing computational cost. The formulation retains robust contact handling via LM while eliminating the need to tune parameters such as contact stiffness and friction, which are difficult to estimate in vivo and are typically set empirically in prior works [20].

Beyond these differences, this work is, to our knowledge, the first to integrate interactive FEM, catheter sensing, and MR for endovascular navigation.

### 4.2 Reliability and Performance

The median *relative displacement error* was below the uncertainty of stereo-based wall reconstruction. Even the maximum 95[th]-percentile error (9.97 mm) remained smaller than lumen diameters in both vascular segments (10.19–15.76 mm femoral; 13.38–20.36 mm vena cava), supporting the suitability of the results of the presented study.

Accurate CT-EM registration is critical because any offset uniformly shifts the reconstructed catheter centerline relative to the vessel wall, producing a comparable shift in predicted contact onset. In our setup, the catheter radius was 3.7 mm, and the smallest vessel radius, located at the femoral insertion cross-section, was 5.095 mm, yielding a minimal clearance of 1.395 mm. A registration error exceeding this clearance would trigger immediate contact at insertion and prevent the simulation from running. The measured

RMSE (1.011 mm) remained below this threshold, preserving the expected contact behavior. Beyond the insertion site, this effect is further mitigated by the larger cross-sections along the vessel anatomy. More generally, assessing applicability requires comparing the catheter–vessel radial clearance to the registration error, as this determines the sensitivity of contact detection to misalignment. In practice, this can be evaluated prior to simulation by inspecting the registered catheter insertion point within the first, and typically most restrictive, vessel cross-section, where the vessel radius reaches its minimum.

The latency of the proposed MARS is dominated by the computational delay of the interactive FEM. During the highest-latency phase (32 s), the FEM accumulated an effective lag of approximately 14 s. Although this prevents strict real-time performance, the update rate remains adequate for interactive use, especially given that catheter–vessel interaction is currently assessed only intermittently via fluoroscopy [24]. Thus, the proposed continuous feedback provides complementary value. GPU-accelerated FEM methods can further reduce delays [25], and slower catheter advancement would also ease computation.

### 4.3 Limitations

Evaluation was limited to one vessel model, though its tortuosity and caliber changes created a challenging case. Displacement measurements from external markers are sensitive to phantom wall-thickness variability, which may affect FEM comparison.

This study used rigid, landmark-based registration, which is unsuitable in a real clinical setting, wherein patient positioning and vessel geometry would likely change from the acquisition of pre-operative CT scans and the procedure. Clinical translation would require deformable registration between the vessel avatar reconstructed from pre-procedural CT scans to the real vessel geometry. To this aim, intraprocedural US imaging could be exploited: the real vessel could be swept at constant speed by a US probe manipulated by a robotic arm, thus acquiring multiple vessel cross sections with a known pose of the US probe, and the intra-procedural vessel geometry could hence be reconstructed. The feasibility of such an approach is supported by prior work [21, 22].

Finally, we focused on displacement analysis; contact forces were not assessed due to their sensitivity to *BBox* mechanical properties.

## 5 Conclusion

This study shows that integrating interactive simulation, catheter shape sensing, and mixed reality can support catheter monitoring during endovascular procedures. Beyond monitoring, MARS is proposed an interpretable interface between the operator and an automated robotic and sensed system, enabling more intuitive guidance and better-informed intraoperative decision-making.

## Supplementary Information

The Supplementary Information includes a PDF with additional methods and virtual benchmarking results (S 1–S 2) and a video (S 3) demonstrating the interactive MR visualization and the simulation used in the experiments.

# Acknowledgements

This work was supported by the European Union's Horizon 2020 research and innovation program, under the project ARTERY, grant agreement No. 101017140.

This work was supported by the MUSA—Multilayered Urban Sustainability Action—project, funded by the European Union—NextGenerationEU, under the National Recovery and Resilience Plan (NRRP) Mission 4 Component 2 Investment Line 1.5: Strengthening of research structures and creation of R&D "innovation ecosystems," set up of "territorial leaders in R&D."

# Statements and Declarations

**Conflict of Interest** The authors declare that they have no conflict of interest.

**Ethical Approval** The CT dataset was provided fully anonymized by the clinical partner within the ARTERY Project. According to the institutional and national regulations, the use of anonymized retrospective data for methodological research did not require additional ethical approval.

**Informed Consent** The clinical partner confirmed that informed consent for data use in research was obtained at the time of data collection.